
\documentstyle{amsppt}
\magnification=1200
\baselineskip=18pt
\nologo
\TagsOnRight
\document
\define \ts{\thinspace}
\define \oa{{\frak o}}
\define \g{{\frak g}}

\def\gl{{\frak {gl}}}

\define \Sym{{\frak S}}
\define \spa{{\frak {sp}}}
\define \U{{\operatorname {U}}}

\define \Y{{\operatorname {Y}}}
\define \C{{\Bbb C}}

\define \si{{\Cal C}}

\define \sgn{\text{{\rm sgn}}}

\define \ot{\otimes}
\define \wh{\widehat}
\define \wt{\widetilde}
\define \qdet{\text{{\rm qdet}\ts}}
\define \sdet{\text{{\rm sdet}\ts}}
\define \End{\text{{\rm End}\ts}}

\define \ra{\rightarrow}

\define \Proof{\noindent {\bf Proof. }}

\heading{\bf NONCOMMUTATIVE SYMMETRIC FUNCTIONS AND} \endheading
\heading{\bf LAPLACE OPERATORS FOR CLASSICAL LIE ALGEBRAS} \endheading
\bigskip
\heading{Alexander Molev}
\endheading
\bigskip
\noindent
Centre for Mathematics and its Applications \newline
School of Mathematical Sciences \newline
Australian National University \newline
Canberra, ACT 0200, Australia \newline
(e-mail:\ molev\@pell.anu.edu.au)

\bigskip
\noindent
{\bf September 1994}


\bigskip
\noindent
{\bf Abstract}\newline
New systems of Laplace (Casimir) operators for the
orthogonal and symplectic Lie algebras are constructed.
The operators are expressed in terms of
paths in graphs related to matrices
formed by the generators of these Lie algebras with the use of
some properties of the noncommutative symmetric functions
associated with a matrix. The decomposition of the
Sklyanin determinant into a product of quasi-determinants
play the main role in the construction. Analogous decomposition for the
quantum determinant provides an alternative proof of the known
construction for the Lie algebra $\gl(N)$.

\newpage
\heading
{\bf 0. Introduction}
\endheading
\

A general theory of noncommutative symmetric functions is
developed in the paper by
Gelfand--Krob--Lascoux--Leclerc--Retakh--Thibon [GKLLRT].
Noncommutative analogues and generalizations of
the classical results of the theory of symmetric
functions are obtained. The background of the
noncommutative theory is the Gelfand--Retakh
quasi-determinants [GR1], [GR2] (see also [KL]).
Various results describing the properties of the
noncommutative symmetric functions
and relations between them are obtained and several
examples of applications of these results for different
kinds of specializations are discussed.
Symmetric functions associated with a matrix whose entries
are elements of a noncommutative ring
is one of these applications.
New systems of generators of the
center of the universal enveloping algebra $\U(\gl(N))$ are
constructed by using the properties of the symmetric functions associated
with
the matrix $E$ formed by the generators of the Lie algebra $\gl(N)$.
The key role in this construction is played by the fact
that the
coefficients of the characteristic Capelli
determinant related to the matrix $E$ are central in the algebra
$\U(\gl(N))$ (see Howe [H], Howe--Umeda [HU]) and that this
determinant is factorized by the quasi-determinants related to
submatrices of $E$ (see Gelfand--Retakh [GR1], [GR2]).

The invariance
of the Capelli determinant can be easily proved by using the properties
of the Yangian $\Y(N)=\Y(\gl(N))$ (see, e.g., Nazarov [N],
Nazarov--Tarasov [NT]);
the latter is a
`quantum' deformation of the universal enveloping algebra for the
polynomial current Lie algebra $\gl(N)[x]$
(see, e.g., Takhtajan--Faddeev [TF], Drinfeld [D]).
Namely, the center of the Yangian is generated by the coefficients
of a formal series called the quantum determinant,
and the Capelli determinant
is the image of the quantum determinant under a natural homomorphism
$\Y(N)\to\U(\gl(N))$.

In this paper we give an alternative proof
of the factorization of the Capelli
determinant by using the decomposition of the quantum determinant
in the algebra of formal power series with coefficients from $\Y(N)$.

Further, we apply this approach to the case of the orthogonal
and symplectic Lie algebras $\oa(N)$ and
$\spa(N)$. Here the Yangian $\Y(N)$ should be
replaced by the Olshanski\u\i\ twisted Yangian $\Y^+(N)$ or $\Y^-(N)$
respectively and the quantum determinant by the Sklyanin
determinant (see [O], [MNO]). In Molev [Mo1] a formula for the Sklyanin
determinant was found and the Capelli-type determinant for the
orthogonal and symplectic Lie algebras was constructed.
As in the
case of $\gl(N)$, this determinant
is a polynomial with coefficients from the center of the
universal enveloping algebra.

It turns out that an analogue of the decomposition of the
quantum determinant takes place for the Sklyanin determinant as well.
Moreover, the image of this decomposition under the natural
homomorphism of the twisted Yangian $\Y^{\pm}(N)$ to the universal
enveloping algebra yields a decomposition of the Capelli-type
determinant into a product of quasi-determinants related to
submatrices of the matrix $F$ formed by the generators
of the Lie algebra $\oa(N)$ or $\spa(N)$.

This allows us using some
results of the paper [GKLLRT] to construct
families of Laplace operators and express them in terms of paths
in graphs related to the matrix $F$. While the invariant
polynomials which correspond to both the Capelli determinant and
the Capelli-type determinant
under the Harish-Chandra
isomorphism are the elementary symmetric functions,
the corresponding polynomials for the Laplace operators of these
families are the complete symmetric functions and
the sums of powers of variables (see [GKLLRT]
for the $\gl(N)$-case).

The approach based on the properties of the Yangians was also used in
[Mo2]
where the quantum Liouville formulae (see [N], [MNO]) were applied
to the calculation of
the Perelomov--Popov invariant polynomials [PP].

I am very grateful to
Grigori Olshanski\u\i\ and Maxim Nazarov for useful
discussions.

\newpage
\heading
{\bf 1. Construction of Laplace operators}\endheading
\

Let us consider the Lie algebra $\gl(N)$ and let $\{E_{ij}\}$, where
$1\leq i,j\leq N$ be
its standard basis. The commutation relations in this basis have the form:
$$
[E_{ij},E_{kl}]=\delta_{kj}E_{il}-\delta_{il}E_{kj}.
$$
For $1\leq m\leq N$ denote by $E^{(m)}$ the
$m\times m$-matrix with the entries $E_{ij}$, where
$i,j=1,\dots,m$, and set $\rho_m:=-m+1$.

Let $\Cal E^{(m)}$ denote the complete oriented graph with
the vertices $\{1,\dots,m\}$, the arrow from $i$ to $j$ is labelled by
the $ij$-th matrix element of the matrix $E^{(m)}+\rho_m$.
Then every path in the graph
defines a monomial in the matrix elements in a natural way.
A path from $i$ to $j$ is called {\it simple} if it does not
pass through the vertices $i$ and $j$ except for the beginning and the end
of
the path.

Using this graph introduce the elements $\Lambda_k^{(m)}$, $S_k^{(m)}$,
$\Psi_k^{(m)}$ and $\Phi_k^{(m)}$ of
the universal enveloping algebra $\U(\gl(N))$ as follows
[GKLLRT]: for $k\geq 1$
\bigskip

$(-1)^{k-1}\Lambda_k^{(m)}$ is the sum of all monomials labelling
simple paths in $\Cal E^{(m)}$ of length $k$ going from $m$ to $m$;
\bigskip

$S_k^{(m)}$ is the sum of all monomials labelling paths in
$\Cal E^{(m)}$ of length $k$ going from $m$ to $m$;
\bigskip

$\Psi_k^{(m)}$ is the sum of all monomials labelling paths in
$\Cal E^{(m)}$ of length $k$ going from $m$ to $m$, the coefficient of
each monomial being the length of the first return to $m$;
\bigskip

$\Phi_k^{(m)}$ is the sum of all monomials labelling paths in
$\Cal E^{(m)}$ of length $k$ going from $m$ to $m$, the coefficient of
each monomial being the ratio of $k$ to the number of returns to $m$.
\bigskip

Denote by $L(\lambda)$, $\lambda=(\lambda_1,\dots,\lambda_N)$, the
highest weight representation of the Lie algebra $\gl(N)$. That is,
$L(\lambda)$ is generated by a nonzero vector $v$ such that
$$
E_{ii}v=\lambda_iv,\quad 1\leq i\leq N,
\qquad\text{and}\qquad E_{ij}v=0,\quad 1\leq i<j\leq N.
$$
For $i=1,\dots,N$ set $l_i:=\lambda_i+\rho_i$.

\proclaim
{{\bf Theorem 1.1} {\rm [GKLLRT, Section 7.5]}} The center of the algebra
$\U(\gl(N))$ is generated by the scalars and each of the
following families of elements.
$$
\align
\Lambda_k=&\sum_{i_1+\cdots+i_N=k}\Lambda_{i_1}^{(1)}\cdots
\Lambda_{i_N}^{(N)},\tag1.1\\
S_k=&\sum_{i_1+\cdots+i_N=k}S_{i_1}^{(1)}\cdots
S_{i_N}^{(N)},\tag1.2\\
\Psi_k=&\sum_{m=1}^N\Psi_k^{(m)},\tag1.3\\
\Phi_k=&\sum_{m=1}^N\Phi_k^{(m)},\tag1.4
\endalign
$$
$k=1,2,\dots,N$. Moreover, $\Psi_k=\Phi_k$ for any $k$, and
the eigenvalues of $\Lambda_k$, $S_k$ and
$\Psi_k$ in the representation $L(\lambda)$ are, respectively,
the elementary, complete and power sums symmetric functions of degree $k$
in the variables $l_1,\dots,l_N$.
\endproclaim

Now we formulate an analogue of this theorem for
the orthogonal and symplectic
Lie algebras. Set $\g(n):=
\oa(2n)$, $\oa(2n+1)$ or $\spa(2n)$. We shall consider
all the three
cases simultaneously. It will be convenient to parametrize
the basis elements $E_{ij}$ of the Lie algebra $\gl(N)$
by the indices $i,j=-n,-n+1,\dots,n-1,n$, where $n:=[N/2]$ and the
index $0$ is skipped when $N$ is even. Then the Lie algebra $\g(n)$
can be realized as a subalgebra in $\gl(N)$ spanned by the
elements
$$
F_{ij}:=\cases E_{ij}-E_{-j,-i},\quad&\text{in the orthogonal case};\\
E_{ij}-\sgn(i)\sgn(j)E_{-j,-i},
\quad&\text{in the symplectic case}.\endcases
$$
\medskip
\noindent
For $i=1,\dots,n$ set
$$
\rho_{-i}=-\rho_i:=\cases i-1\quad&\text{for}\quad\g(n)=\oa(2n),\\
i-\frac12\quad&\text{for}\quad\g(n)=\oa(2n+1),\\
i\quad&\text{for}\quad\g(n)=\spa(2n).
\endcases
$$
We also set $\rho_0:=1/2$ in the case
of $\g(n)=\oa(2n+1)$.
Let $1\leq m\leq n$.
Denote by $F^{(m)}$ the matrix with the entries $F_{ij}$, where
$i,j=-m,-m+1,\dots,m$ (the index $0$ is skipped if $N=2n$).
Let us consider the complete oriented graph $\Cal F_m$ with
the vertices $\{-m,-m+1,\dots,m\}$, the arrow from $i$ to $j$ is labelled
by
the $ij$-th matrix element of the matrix $F^{(m)}+\rho_m$.
\medskip

Introduce now the elements $\Lambda_{k}^{(m)}$, $\wt{\Lambda}_{k}^{(m)}$,
$S_{k}^{(m)}$, $\wt S_{k}^{(m)}$,
$\Phi_{k}^{(m)}$, $\wt{\Phi}_{k}^{(m)}$ of
the universal enveloping algebra $\U(\g(n))$ as follows: for $k\geq 1$
\bigskip

$(-1)^{k-1}\Lambda_{k}^{(m)}$ (resp. $-\wt{\Lambda}_{k}^{(m)}$)
is the sum of all monomials labelling
simple paths in $\Cal F^{(m)}$
(resp. simple paths that do not pass through $-m$)
of length $k$ going from $m$ to $m$;
\bigskip

$S_{k}^{(m)}$ (resp. $(-1)^k\wt S_{k}^{(m)}$)
is the sum of all monomials labelling paths in
$\Cal F^{(m)}$ (resp. paths that do not pass through $-m$)
of length $k$ going from $m$ to $m$;
\bigskip

$\Phi_{k}^{(m)}$ (resp. $(-1)^k\wt{\Phi}_{k}^{(m)}$)
is the sum of all monomials labelling paths in
$\Cal F^{(m)}$ (resp. paths that do not pass through $-m$)
of length $k$ going from $m$ to $m$, the coefficient of
each monomial being the ratio of $k$ to the number of returns to $m$.
\bigskip

Denote by $L(\lambda)$, $\lambda=(\lambda_1,\dots,\lambda_n)$, the
highest weight representation of the Lie algebra $\g(n)$. That is,
$L(\lambda)$ is generated by a nonzero vector $v$ such that
$$
F_{ii}v=\lambda_iv,\quad 1\leq i\leq n;
\qquad\text{and}\qquad F_{ij}v=0,\quad -n\leq i<j\leq n.
$$
For $i=1,\dots,n$ set $l_i:=\lambda_i+\rho_i$.

\proclaim
{\bf Theorem 1.2} Each of the
following families of elements is contained in the center of the algebra
$\U(\g(n))$:
$$
\align
\Lambda_{2k}=&\sum_{i_1+\cdots+i_{2n}=2k}\wt{\Lambda}_{i_1}^{(1)}
\Lambda_{i_2}^{(1)}\cdots \wt{\Lambda}_{i_{2n-1}}^{(n)}
\Lambda_{i_{2n}}^{(n)},\tag1.5\\
S_{2k}=&\sum_{i_1+\cdots+i_{2n}=2k}\wt{S}_{i_1}^{(1)}
S_{i_2}^{(1)}\cdots \wt{S}_{i_{2n-1}}^{(n)}
S_{i_{2n}}^{(n)},\tag1.6\\
\Phi_{2k}=&\sum_{m=1}^n(\wt{\Phi}_{2k}^{(m)}
+{\Phi}_{2k}^{(m)}),\tag1.7
\endalign
$$
$k=1,2,\dots$\ts. Moreover, the eigenvalues of
$(-1)^k\Lambda_{2k}$, $S_{2k}$ and
$\Phi_{2k}/2$ in the representation $L(\lambda)$ are, respectively,
the elementary, complete and power sums symmetric functions of degree $k$
in the variables $l_1^2,\dots,l_n^2$.
\endproclaim

\bigskip
\noindent
{\bf Remarks.} (i) One can define
the elements $\Psi_{2k}$
by analogy with the $\gl(N)$-case. However, the assertion
of Theorem 1.2 for them is wrong.

(ii) Replacing $2k$ with $2k-1$ in the formulae (1.5)--(1.7)
one could define the central elements $\Lambda_{2k-1}$, $S_{2k-1}$ and
$\Phi_{2k-1}$, but they turn out to be equal to $0$.

(iii) The scalars and each of the families (1.5)--(1.7) with $k=1,\dots,n$
generate the center of $\U(\g(n))$, except for the case of $\g(n)=\oa(2n)$.
To get generators
in the latter case, the $2n$-th elements in each family
should be replaced with
$\Lambda_{2n}^{1/2}$ which coincides,
up to a constant, with the central Pfaffian-type element; see [Mo1].

\newpage
\heading
{\bf 2. Quasi-determinants and noncommutative symmetric
functions}\endheading
\

The main point in the proof of Theorems 1.1 and 1.2 is the use of
the decomposition of the Capelli-type determinants into a product of
quasi-determinants. In the case of the Lie algebra $\gl(N)$
this decomposition was obtained in [GR1] and used in
[GKLLRT, Section 7.4] to prove Theorem 1.1.
We outline that proof here. Then we formulate
an analogue of this decomposition for
the orthogonal and symplectic Lie algebras (for the proof see Section 4)
and using some ideas
of the paper [GKLLRT] derive Theorem 1.2.
\medskip

The {\it algebra of noncommutative symmetric functions\/} [GKLLRT]
is defined as the free associative algebra with countably many
generators $\Lambda_1$,\ts$\Lambda_2,\dots\ts$\ts. These generators are
called
the {\it elementary symmetric functions}.
The {\it complete symmetric functions} $S_k$,
the {\it power sums symmetric functions of the first kind} $\Psi_k$
and
the {\it power sums symmetric functions of the second kind} $\Phi_k$
are defined as follows. Set
$$
\lambda(t)=1+\sum_{k=1}^{\infty}\Lambda_kt^k,\tag2.1
$$
where $t$ is an intermediate commuting with all the $\Lambda_k$. Then
$$
\align
1+\sum_{k=1}^{\infty}S_kt^k&:=\lambda(-t)^{-1},\tag2.2\\
\sum_{k=1}^{\infty}\Psi_kt^{k-1}&:=\lambda(-t)
\cdot {d\over dt}(\lambda(-t)^{-1}),\tag2.3\\
\sum_{k=1}^{\infty}\Phi_kt^{k-1}&:=-{d\over dt}\log \lambda(-t).\tag2.4
\endalign
$$
These relations generalize the corresponding relations between
classical symmetric functions; see, e.g. Macdonald [M].  In particular,
in the commutative case
formulae (2.3) and (2.4) give the same families of
power sums functions, that is,
$\Psi_k=\Phi_k$ for all $k=1,2,\dots\ts$.
Various applications of both the commutative and
noncommutative theory can be
obtained by means of specializations. That is, the series $\lambda(t)$
is replaced with an arbitrary series whose coefficients are
still regarded as (formal) elementary symmetric functions. Then
the other symmetric functions are constructed by
using formulae (2.2)--(2.4)\newline
(see [M], [GKLLRT]).
\medskip

Let $X$ be an $n\times n$-matrix over a ring with the unity
and suppose that
there exists the matrix $X^{-1}$, and its $ji$-th entry $(X^{-1})_{ji}$ is
invertible. Then the $ij$-{\it th quasi-determinant of} $X$ is defined
by the formula
$$
|X|_{ij}:=\bigl((X^{-1})_{ji}\bigr)^{-1}.
$$
Note that the definition of $|X|_{ij}$ can be made much less restrictive.
In particular, the quasi-determinant can be defined for some
non-invertible matrices (see [GR1], [GR2]). If the ring is commutative,
then $|X|_{ij}=(-1)^{i+j}\det X/\det X^{ij}$
where $X^{ij}$ is the submatrix of $X$ obtained by
removing the $i$-th row and $j$-th column.
\medskip

Let $E$ denote the $N\times N$-matrix with the entries $E_{ij}$
and $\rho$ denote the diagonal
matrix with the diagonal entries $\rho_1,\dots,\rho_N$.
Consider the {\it Capelli determinant\/}
$\det\ts(1+t(E+\rho))$, where `det' denotes the usual alternating sum of
the
products of the entries of the matrix, provided that the first
element in each product is taken from the first column, the second one
from the second column etc. This determinant is a polynomial in $t$
of degree $N$ and it was proved in Howe--Umeda [HU] that all its
coefficients belong to the center of the algebra $\U(\gl(N))$.
On the other hand,
one has the following
decomposition of the Capelli determinant into the product of
quasi-determinants.

\proclaim {{\bf Theorem 2.1} {\rm [GR1]}} In the algebra of
formal series in $t$ with coefficients from $\U(\gl(N))$ one has
$$
\det\ts(1+t(E+\rho))=(1+tE_{11})\ts|1+t(E^{(2)}+\rho_2)|_{22}\cdots
|1+t(E^{(N)}+\rho_N)|_{NN}.\tag2.5
$$
\endproclaim
\medskip

Now we apply this decomposition to prove Theorem 1.1.
It was proved in [GKLLRT, Proposition 7.20] that the generating series
of the elements $\Lambda_k^{(m)}$, $S_k^{(m)}$,
$\Psi_k^{(m)}$ and $\Phi_k^{(m)}$ introduced in Section 1 can be written
in terms of quasi-determinants in the following way:
$$
1+\sum_{k=1}^{\infty}\Lambda_k^{(m)}t^k=|1+t(E^{(m)}+\rho_m)|_{mm},\tag2.6
$$
$$
1+\sum_{k=1}^{\infty}S_k^{(m)}t^k=|1-t(E^{(m)}+\rho_m)|_{mm}^{-1},\tag2.7
$$
$$
\sum_{k=1}^{\infty}\Psi_k^{(m)}t^{k-1}
=|1-t(E^{(m)}+\rho_m)|_{mm}\ts
{d\over dt}\ts|1-t(E^{(m)}+\rho_m)|_{mm}^{-1},\tag2.8
$$
$$
\sum_{k=1}^{\infty}\Phi_k^{(m)}t^{k-1}
=-{d\over dt}\log(|1-t(E^{(m)}+\rho_m)|_{mm}).\tag2.9
$$
Applying Theorem 2.1 to the Lie algebras $\gl(m)$ where
$m=1,\dots,N$ we see that the quasi-determinant
$|1+t(E^{(m)}+\rho_m)|_{mm}$
is represented as the ratio of two Capelli determinants corresponding
to the Lie algebras $\gl(m)$ and $\gl(m-1)$. This implies that the
elements $\Lambda_k^{(m)}$ with $k\geq 1$ and $1\leq m\leq N$
generate a commutative subalgebra in $\U(\gl(N))$
[GKLLRT, Theorem 7.26].

On the other hand, formulae (2.5) and (2.6) imply that the coefficients of
the Capelli determinant coincide with the elements $\Lambda_k$ given by
(1.1);
that is,
$$
1+\Lambda_1t+\cdots+\Lambda_Nt^N=\det\ts(1+t(E+\rho)).\tag2.10
$$
Let us regard now these elements as specializations of
(commutative) elementary symmetric functions. Then, comparing
formulae (2.2)--(2.4) with (2.5)--(2.9) we conclude that
the corresponding specializations of the
complete symmetric functions $S_k$ and the power sums
symmetric functions $\Psi_k(=\Phi_k)$ are given by formulae (1.2)--(1.4).
This proves the first part of Theorem 1.1.

It is clear that to prove the second part it suffices to
find the eigenvalues of the elements of one of the families (1.1)--(1.4)
in the representation $L(\lambda)$. It can be easily done for
the coefficients of the Capelli determinant as well as for
the power sums functions $\Psi_k$ (or $\Phi_k$). The proof is complete.

\bigskip
\noindent
{\bf Remark.} It can be easily seen from formulae (2.1)--(2.4) and
(2.6)--(2.9) that
for any fixed $m$ the elements $\Lambda_k^{(m)}$,
$S_k^{(m)}$, $\Psi_k^{(m)}$ and $\Phi_k^{(m)}$ can be regarded as
specializations of the corresponding noncommutative
symmetric functions.
These elements were introduced in [GKLLRT] as a special case of the
{\it noncommutative
symmetric functions associated with a matrix\/} over an arbitrary
ring.
However, as we have seen in the above proof,
these functions (associated with the matrix $E^{(m)}+\rho_m$)
form a {\it commutative\/} subalgebra in
$\U(\gl(N))$.

\medskip

Let us turn now to the case of the orthogonal and symplectic
Lie algebras. We shall keep using the notation introduced in Section 1.

\medskip

Let $F$ denote the $N\times N$-matrix with the entries $F_{ij}$, where
$-n\leq i,j\leq n$ and let $\rho$ denote the diagonal matrix with the
diagonal
entries $\rho_{-n},\dots,\rho_n$.
We introduce an analogue of the Capelli determinant
for the orthogonal and symplectic Lie algebras in the following way.
First, we define a map
$$
\Sym_N\ra \Sym_{N-1},\qquad p\mapsto p' \tag 2.11
$$
of the symmetric group $\Sym_N$ to its natural subgroup $\Sym_{N-1}$
as follows (see [Mo1, Section 3]).
Let us consider the graph $\Gamma_N$ whose vertices are
identified with the elements
$\Sym_N$, and two permutations $p=(p(1),\dots,p(N))$
and $q=(q(1),\dots,q(N))$ are connected by an edge if $q$ can be
obtained from $p$ by interchanging two indices
$p(k)$ and $p(l)$, provided that
either
\medskip

(i) $p(k)$ and $p(l)$ are the two maximal elements of the set
$\{p(i),p(i+1),\dots,p(N-i+1)\}$ for some $1\leq i\leq n$; or
\medskip

(ii) $k<l<N-k+1$ and $p(k),p(l),p(N-k+1)$ are the three maximal
elements of the set $\{p(k),p(k+1),\dots,p(N-k+1)\}$.
\medskip

The graph $\Gamma_N$ admits the following properties. It has $(N-1)!$
connected components and each of them is isomorphic (as a graph) to
the $k$-dimensional cube for some $1\leq k\leq N-1$. Moreover,
given $k$ the number of components, isomorphic to
the $k$-dimensional cube, equals the {\it signless Stirling number of the
first kind}
$c\thinspace(N-1,k)$ (see, e.g., Stanley [S]).
Each connected component of the graph $\Gamma_N$ contains
a unique vertex $p_0=(p_0(1),\dots,p_0(N))$ such that $p_0(n+1)=N$.
Then, by definition, all
the vertices of this component have the same image
$p'=(p_0(N),\dots,p_0(n+2),p_0(n),\dots,p_0(1))$ under the map
(2.11).

Now the {\it Capelli-type determinant\/} $\wt{\det}\ts(1+t(F+\rho))$
is defined by the formula
$$
\aligned
\wt{\det}\ts(1+t(F+\rho))=&\\
(-1)^n\sum_{p\in\Sym_N}&\sgn(pp')
(1+t(F+\rho_{-n}))_{-i_{p(1)},i_{p'(1)}}
\cdots(1+t(F+\rho_{n}))_{-i_{p(N)},i_{p'(N)}},
\endaligned\tag 2.12
$$
where $(i_1,\dots,i_N)$ is
an arbitrary permutation of the indices $(-n,\dots,n)$ (the right hand side
does not depend on this permutation).

It was proved in [Mo1] that this determinant can be represented in the form
$$
\wt{\det}\ts(1+t(F+\rho))=\Lambda(t^2),
\quad\text{if}\quad N=2n\quad\text{and}\quad
$$
$$
\wt{\det}\ts(1+t(F+\rho))=(1+t/2)\ts \Lambda(t^2),
\quad\text{if}\quad N=2n+1,
$$
where $\Lambda(s)$ is a polynomial in $s$ of degree $n$ whose coefficients
belong to the center of the algebra
$\U(\g(n))$.

For $1\leq m\leq n$ denote by $\wt F^{(m)}$ the submatrix of the
matrix $F^{(m)}$ (see Section 1) obtained by removing the row and column
enumerated by $-m$.
The following analogue of Theorem 2.1 will be proved in Section 4.

\proclaim {\bf Theorem 2.2} In the algebra of
formal series in $t$ with coefficients from $\U(\g(n))$ one has
the decomposition:
$$
\Lambda(t^2)=|1-t(\wt F^{(1)}+\rho_1)|_{11}\cdot
|1+t(F^{(1)}+\rho_1)|_{11}
\cdots|1-t(\wt F^{(n)}+\rho_n)|_{nn}\cdot
|1+t(F^{(n)}+\rho_n)|_{nn}.\tag2.13
$$
Moreover, all the factors on the right
hand side of {\rm (2.13)} are permutable.
\endproclaim

Using this result we prove now Theorem 1.2.
We have the following generating series for the elements
$\Lambda_{k}^{(m)}$,
$S_{k}^{(m)}$ and
$\Phi_{k}^{(m)}$
introduced in Section 1 (see [GKLLRT, Proposition 7.20]):
$$
1+\sum_{k=1}^{\infty}\Lambda_k^{(m)}t^k=|1+t(F^{(m)}+\rho_m)|_{mm},\tag2.14
$$
$$
1+\sum_{k=1}^{\infty}S_k^{(m)}t^k=|1-t(F^{(m)}+\rho_m)|_{mm}^{-1},\tag2.15
$$
$$
\sum_{k=1}^{\infty}\Phi_k^{(m)}t^{k-1}
=-{d\over dt}\log(|1-t(F^{(m)}+\rho_m)|_{mm}).\tag2.16
$$
Similarly, for
the generating series of the elements
$\wt{\Lambda}_{k}^{(m)}$,
$\wt S_{k}^{(m)}$,
$\wt{\Phi}_{k}^{(m)}$ we have
$$
1+\sum_{k=1}^{\infty}\wt{\Lambda}_k^{(m)}t^k=
|1-t(\wt F^{(m)}+\rho_m)|_{mm},\tag2.17
$$
$$
1+\sum_{k=1}^{\infty}\wt S_k^{(m)}t^k=
|1+t(\wt F^{(m)}+\rho_m)|_{mm}^{-1},\tag2.18
$$
$$
\sum_{k=1}^{\infty}\wt{\Phi}_k^{(m)}t^{k-1}
=-{d\over dt}\log(|1+t(\wt F^{(m)}+\rho_m)|_{mm}).\tag2.19
$$
We obtain from formulae (2.13), (2.14)
and (2.17) that the coefficients of the
polynomial $\Lambda(s)$ coincide with the elements $\Lambda_{2k}$
given by (1.5); that is,
$$
1+\Lambda_2t^2+\cdots+\Lambda_{2n}t^{2n}=\Lambda(t^2).
$$
As in the $\gl(N)$-case, let us regard
this polynomial (in the variable $t$) as
a specialization of the generating series $\lambda(t)$
for the elementary symmetric functions (see (2.1)). Then using
Theorem 2.2 and formulae (2.14)--(2.19) we obtain that the
corresponding specializations of
the complete symmetric functions $S_{2k}$ and
the power sums symmetric functions $\Phi_{2k}$ are given by (1.6) and
(1.7).
This proves the first part of Theorem 1.2.

To prove the second part we can either
find the eigenvalues of the elements $\Phi_{2k}$ in the
representation $L(\lambda)$ directly, or use the fact (see [Mo1]) that
the image of the polynomial $\Lambda(s)$ in $L(\lambda)$ is
$(1-sl_1^2)\cdots (1-sl_n^2)$, which completes the proof of Theorem 1.2.

\newpage
\heading
{\bf 3. Yangian for $\gl(N)$ and the decomposition}
\endheading
\heading
{\bf of the quantum determinant}
\endheading
\

Here we use the approach based on the properties of the Yangians
to obtain an alternative proof of the decomposition
of the Capelli determinant into the product of
quasi-determinants (Theorem 2.1). Various results describing the algebraic
structure of the Yangians are collected in the paper [MNO]. We
reproduce those of them we shall use below.
\medskip

The {\it Yangian} $\Y(N)=\Y(\gl(N))$ is the
complex associative algebra with the
generators $t_{ij}^{(1)},t_{ij}^{(2)},\dots$ where $1\leq i,j\leq N$.
The quadratic defining relations are written in the
following way. First,
for any $i,j=1,\ldots, N$ introduce the formal power series
$$
t_{ij}(u) = \delta_{ij} + t^{(1)}_{ij} u^{-1} + t^{(2)}_{ij}u^{-2} +
\cdots \in \Y(N)[[u^{-1}]] \tag3.1
$$
and combine these series into a single $T$-{\it matrix}:
$$
T(u):=\sum^N_{i,j=1} t_{ij}(u)\ot E_{ij}
\in \Y(N)[[u^{-1}]]\ot \End(\C^N), \tag3.2
$$
where $E_{ij}$ are the standard matrix units.
We need to consider the multiple tensor products of the form
$\C^N\ot\dots\ot\C^N$
and operators
therein.
For an operator $X\in \End(\C^N)$ and a number $m=1,2,\ldots$ we set
$$
X_k:=1^{\otimes (k-1)}\otimes X\otimes 1^{\otimes (m-k)}
\in\End(\C^N)^{\otimes m}, \quad 1\leq k\leq m. \tag3.3
$$
If $X\in\End(\C^N)^{\otimes 2}$ then for any $k, l$\  such that
$1\leq k, l\leq m$\ and $k\neq l$, we denote by $X_{kl}$ the operator in
$(\C^N)^{\otimes m}$
which acts as $X$ in the product of $k$-th and $l$-th copies and as
1 in all other copies. That is,
$$
\aligned
&X=\sum_{r,s,t,u} a_{rstu} E_{rs}\otimes E_{tu}, \quad
a_{rstu}\in\C\qquad \Rightarrow\\
\Rightarrow \quad &X_{kl}=\sum_{r,s,t,u} a_{rstu} (E_{rs})_k\ (E_{tu})_l.
\endaligned\tag3.4
$$
Further, given formal variables $u_1,\dots,u_m$ we set for $k=1,\dots,m$
$$
T_k (u_k) :=\sum_{i,j=1}^N t_{ij} (u_k) \otimes (E_{ij})_k
\in \Y(N)[[u_1^{-1},\ldots,u_m^{-1}]]\otimes\End(\C^N)^{\otimes m}.
\tag 3.5
$$
We let $P$ denote the permutation operator in $\C^N\ot \C^N$:
$$
P:=\sum_{i,j}E_{ij}\ot E_{ji}.
$$
Now the defining
relations for $t_{ij}^{(k)}$ can be written as the {\it ternary relation}
on
the $T$-matrix:
$$
R(u-v)T_1(u)T_2 (v) = T_2(v)T_1(u)R(u-v),   \tag 3.6
$$
where
$$
R(u)=R_{12}(u):=1-u^{-1}P_{12}
$$
is the {\it Yang} $R$-{\it matrix}.
\medskip

Let
$$
a_N=(N!)^{-1}\sum_{p\in \Sym_N}\sgn(p)\cdot p\in\C[\Sym_N]
$$
denote the normalized antisymmetrizer in the group ring. Consider the
natural
action of $\Sym_N$ in the tensor space $(\C^N)^{\otimes N}$
and denote by $A_N$ the image of $a_N$.

There exists a formal series
$$
\qdet T(u):=1+d_1u^{-1}+d_2u^{-2}+\cdots \in \Y(N)[[u^{-1}]] \tag 3.7
$$
such that the following identity holds:
$$
A_NT_1(u)\cdots T_N(u-N+1)=\qdet T(u)A_N.
\tag3.8
$$
The series $\qdet T(u)$ is called the {\it quantum determinant}
of the matrix $T(u)$. Explicit formulae for the quantum determinant
can be easily derived from the identity (3.8).
We shall use the following one below:
$$
\qdet T(u)=\sum_{p\in \Sym_N} \sgn(p)\ts t_{p(1),1}(u)\cdots
t_{p(N),N}(u-N+1).\tag3.9
$$
The coefficients $d_1,d_2,\dots$ of the quantum
determinant $\qdet T(u)$ are algebraically independent generators of
the center of the algebra $\Y(N)$.

For $1\leq m\leq N$ denote by $T^{(m)}(u)$
the submatrix of $T(u)$ corresponding the
first $m$ rows and columns.

\proclaim {\bf Theorem 3.1} One has the following decomposition
of $\qdet T(u)$ in the algebra\newline $\Y(N)[[u^{-1}]]$:
$$
\qdet T(u)=t_{11}(u)\ts|T^{(2)}(u-1)|_{22}\cdots
|T^{(N)}(u-N+1)|_{NN}.\tag3.10
$$
\endproclaim

\Proof In fact, this theorem is a special case of Theorem 7.3 from [MNO]
(see also [KS], [NT]),
and easily follows from formula (3.8). Indeed,
let us define the matrix $\wh T(u)=(\wh t_{ij}(u))$
by the formula
$$
\wh T(u)=\qdet T(u)\ts T^{-1}(u-N+1).\tag3.11
$$
Then multiplying both sides of (3.8) by $T_N^{-1}(u-N+1)$ from the right
we obtain the relation
$$
A_NT_1(u)\cdots T_{N-1}(u-N+2)=A_N\wh T_N(u). \tag3.12
$$
Taking the $NN$-th entry of the matrices
on the left and right hand sides of (3.11) we get
$$
\qdet T(u)\ts(T^{-1}(u-N+1))_{NN}=\wh t_{NN}(u).
$$
It follows from (3.12) that $\wh t_{NN}(u)=\qdet T^{(N-1)}(u)$,
and so,
$$
\qdet T(u)=\qdet T^{(N-1)}(u)\ts|T(u-N+1)|_{NN}.
$$
An easy induction proves the theorem.
\medskip

Let us show now that the decomposition (2.5) is a consequence of (3.10).
Indeed, it is not difficult to verify [D] (see also [MNO]) that
the mapping
$$
\xi : t_{ij}(u)\mapsto \delta_{ij}+E_{ij}u^{-1}
$$
defines the algebra homomorphism
$$
\xi : \Y(N)\ra \U(\gl(N)).
$$
Formula (3.9)
implies that
$$
\prod_{m=1}^N(1+\rho_mu^{-1})\ts\xi(\qdet T(u))=
\det\ts(1+(E+\rho)u^{-1}),
$$
while
$$
(1+\rho_mu^{-1})\ts\xi(|T^{(m)}(u-m+1)|_{mm})=
|1+(E^{(m)}+\rho_m)u^{-1}|_{mm}.
$$
So, applying the homomorphism $\xi$ to both sides of (3.10)
and replacing $u^{-1}$ with $t$, we obtain
the decomposition (2.5) of the Capelli determinant.

\newpage
\heading
{\bf 4. Twisted Yangian and the decomposition}\endheading
\heading{\bf of the Sklyanin determinant}
\endheading
\

In this section following Olshanski\u\i\ [O], [MNO] we define
the algebra $\Y^{\pm}(N)$ which is an analogue
of the Yangian for the orthogonal and symplectic Lie algebras
and introduce a formal series --- the Sklyanin
determinant whose coefficients generate the center of $\Y^{\pm}(N)$. Then
we obtain an analogue of Theorem 3.1 for the Sklyanin
determinant and use it to prove Theorem 2.2.
\medskip

Let $\{e_i\}$
be the canonical basis of the vector space $\C^N$.
{}From now on we shall parametrize
the basis vectors by the numbers $i=-n,-n+1,\ldots,n-1,n$, where $n:=[N/2]$
and $i=0$ is skipped when $N$ is even.

It will be convenient to use
the symbol $\theta_{ij}$ which is defined as follows:
\medskip
$$
\theta_{ij}:=\cases 1,\quad&\text{in the orthogonal case};\\
\sgn(i)\sgn(j),\quad&\text{in the symplectic case}.\endcases
$$
\bigskip

Whenever the double sign $\pm{}$ or $\mp{}$ occurs,
the upper sign corresponds to the orthogonal case and the lower sign to
the symplectic one.

By $X\mapsto X^t$ we will denote the matrix transposition
which is defined on the matrix units as follows:
$$
(E_{ij})^t=\theta_{ij}E_{-j,-i}.
$$
\medskip

The {\it twisted Yangian} $\Y^{\pm}(N)$ is the complex associative algebra
with the generators
$s_{ij}^{(1)},s_{ij}^{(2)},\dots$, where $-n\leq i,j\leq n$. The defining
relations are written in the following way. First, for $-n\leq i,j\leq n$
introduce the formal power series
$$
s_{ij}(u)=\delta_{ij}+s_{ij}^{(1)}u^{-1}+s_{ij}^{(2)}u^{-2}+\cdots\tag4.1
$$
and set
$$
S(u)=\sum_{i,j} s_{ij}(u)\ot
E_{ij}\in\Y^{\pm}(N)[[u^{-1}]]\ot\End(\C^N).\tag4.2
$$
Further, introduce the following element of the algebra
$\End(\C^N)^{\ot2}$:
$$
Q:=\sum_{i,j}\theta_{ij}E_{-j,-i}\ot E_{ji},
$$
and set
$$
R'(u)=1-u^{-1}Q.
$$
As in the case of the Yangian $\Y(N)$ we adopt here the
notation of type (3.3)--(3.5) for the twisted Yangian.
Now the defining relations for $s_{ij}^{(k)}$ can be written
as the {\it quaternary relation} and
the {\it symmetry relation} for the matrix $S(u)$:
$$
R(u-v)S_1(u)R'(-u-v)S_2(v)=S_2(v)R'(-u-v)S_1(u)R(u-v), \tag 4.3
$$
$$
S^t(-u)=S(u)\pm {S(u)-S(-u)\over 2u}.\tag 4.4
$$

Note that the twisted Yangian $\Y^{\pm}(N)$ can be identified with a
subalgebra in $\Y(N)$ by setting $S(u)=T(u)T^t(-u)$.

There exists a formal series
$$
\sdet S(u)\in \Y^{\pm}(N)[[u^{-1}]]
$$
such that the following identity holds:
$$
\align
A_NS_1(u)R'_{12}\cdots &R'_{1N}S_2(u-1)
R'_{23}\cdots R'_{2N}S_3(u-2)\\
\cdots &S_{N-1}(u-N+2)R'_{N-1,N}S_N(u-N+1)=\sdet S(u)A_N,\tag4.5
\endalign
$$
where $R'_{ij}:=R'_{ij}(-2u+i+j-2)$.
The series $\sdet S(u)$ is called
the {\it Sklyanin determinant} of the matrix $S(u)$.
A formula for $\sdet S(u)$ was obtained in [Mo1]. To write it down we
use the map (2.11). For any permutation $(i_1,\dots,i_N)$ of the
indices $(-n,\dots,n)$ we have
$$
\aligned
\sdet S(u)=(-1)^n\gamma_N(u)\sum_{p\in\Sym_N}\sgn(pp')
s^t_{-i_{p(1)},i_{p'(1)}}(-u)\cdots
s^t_{-i_{p(n)},i_{p'(n)}}(-u+n-1)&\\
\cdot\ts s_{-i_{p(n+1)},i_{p'(n+1)}}(u-n)\cdots
s_{-i_{p(N)},i_{p'(N)}}(u-N+1)&,
\endaligned\tag4.6
$$
\medskip
\noindent
where $s^t_{ij}(u)$ are the matrix elements of the matrix $S^t(u)$ and
$\gamma_N(u)\equiv 1$ in the orthogonal case and
$\gamma_N(u)=(2u+1)/(2u-N+1)$ in the symplectic case.

Let us set
$$
c(u):={1\over\gamma_N(u+N/2-1/2)}\ts
\sdet S(u+N/2-1/2).
$$
\medskip
\noindent
Then $c(u)$ is an even formal series in $u^{-1}$,
$c(u)=1+c_2u^{-2}+c_4u^{-4}+\cdots$,
and the elements $c_2,c_4,\dots$ are algebraically
independent generators of the center of the algebra $\Y^{\pm}(N)$.

Let $1\leq m\leq n$.
Denote by $S^{(m)}(u)$ the submatrix of $S(u)$
corresponding the rows and columns enumerated by $-m,-m+1,\dots,m$,
and by $\wt S^{(m)}(u)$ the submatrix of $S^{(m)}(u)$ obtained by removing
the row and column enumerated by $-m$. We have the following analogue of
Theorem 3.1.

\proclaim
{\bf Theorem 4.1} If $N=2n$ then
$$
\align
c(u)=|\wt S^{(1)}(-u-1/2)|_{11}\cdot&\ts|S^{(1)}(u-1/2)|_{11}\\
\cdots &\ts|\wt S^{(n)}(-u-n+1/2)|_{nn}\cdot|S^{(n)}(u-n+1/2)|_{nn},\tag4.7
\endalign
$$
if $N=2n+1$ then
$$
c(u)=s_{00}(u)\cdot|\wt S^{(1)}(-u-1)|_{11}\cdot|S^{(1)}(u-1)|_{11}
\cdots |\wt S^{(n)}(-u-n)|_{nn}\cdot|S^{(n)}(u-n)|_{nn}.\tag4.8
$$
Moreover, all the factors on the right sides of {\rm (4.7)} and
{\rm (4.8)} are permutable.
\endproclaim

\Proof Let us define the matrix $\wh S(u)=
(\wh s_{ij}(u))$ by the formula
$$
\wh S(u)=\sdet S(u)\ts S^{-1}(u-N+1).\tag4.9
$$
Then, multiplying both sides of (4.5) by $S_N^{-1}(u-N+1)$ from the right
we obtain the relation
$$
\align
A_NS_1(u)R'_{12}\cdots R'_{1N}S_2(u-1)&
R'_{23}\cdots R'_{2N}S_3(u-2)\\
\cdots \ts &S_{N-1}(u-N+2)R'_{N-1,N}
=A_N\wh S_N(u).\tag4.10
\endalign
$$
Taking the $nn$-th entry of the matrices on the left and right
hand sides of (4.9) we get
$$
\sdet S(u)\ts (S^{-1}(u-N+1))_{nn}=\wh s_{nn}(u).
$$
Hence,
$$
\sdet S(u)=\wh s_{nn}(u)\ts |S(u-N+1)|_{nn}.\tag4.11
$$

Now we use formula (4.10). It can be easily verified by
using the symmetry relation (4.4) (see [Mo1]) that
$$
A_NS_1(u)R'_{12}\cdots R'_{1N}=\frac{2u+1}{2u\pm 1}\ts
A_NS_1^t(-u).\tag4.12
$$
Denote by $A_N^{(2)}$ the normalized antisymmetrizer corresponding to
the subgroup of $\Sym_N$ consisting of the permutations which
preserve the first index. Clearly, $A_N=A_NA_N^{(2)}$. Note
that $A_N^{(2)}$ is permutable with $S_1^t(-u)$, and $R'_{ij}$
is permutable with
$R'_{kl}$ and $S_k(u)$, provided that the indices $i,j,k,l$ are distinct.
So, we can rewrite formula (4.10) in the form:
$$
\align
\frac{2u+1}{2u\pm 1}\ts A_NS_1^t(-u)A_N^{(2)}&S_2(u-1)
R'_{23}\cdots R'_{2,N-1}S_3(u-2)\\
\cdots \ts &S_{N-1}(u-N+2)R'_{2N}\cdots R'_{N-1,N}
=A_N\wh S_N(u).\tag4.13
\endalign
$$
Let us apply the operators in
both sides of this formula to the vector
$v_i=e_{-i}\ot e_{-n+1}\ot e_{-n+2}\ot\cdots\ot e_{n-1}\ot e_n$, where
$i\in\{-n+1,\dots,n\}$. For the right hand side we clearly obtain
$$
A_N\wh S_N(u)v_i=
\delta_{in}\ts\wh s_{nn}(u)\ts\zeta,
\tag4.14
$$
where $\zeta:=A_N(e_{-n}\ot e_{-n+1}
\ot\cdots\ot e_n)$.
To calculate the left hand side we note first that
$$
R'_{2N}\cdots R'_{N-1,N}v_i=v_i.
$$
Further, let us introduce the formal series
$$
\si_{a_2,\dots,a_{N-1}}(u-1)\in\Y^{\pm}(N)[[u^{-1}]],\quad
-n\leq a_i\leq n,
$$
as follows:
$$
A_N^{(2)}S_2(u-1)
R'_{23}\cdots R'_{2,N-1}S_3(u-2)
\cdots S_{N-1}(u-N+2)(e_{-n+1}\ot\cdots\ot e_{n-1})
$$
$$
=\sum_{a_2,\dots,a_{N-1}}\si_{a_2,\dots,a_{N-1}}(u-1)
(e_{a_2}\ot\cdots\ot e_{a_{N-1}}).
$$
In particular,
$$
(N-2)!\ts\si_{-n+1,\dots,n-1}(u-1)=\sdet S^{(n-1)}(u-1),\tag 4.15
$$
and the series $\si_{a_2,\dots,a_{N-1}}(u-1)$
is skew symmetric with respect to permutations of the indices
$a_2,\dots,a_{N-1}$. This allows us to write the left hand side of
(4.13) applied to $v_i$ in the form:
$$
\frac{2u+1}{2u\pm 1}\ts(N-2)!\sum_{k=1}^{N-1}(-1)^{k-1}\ts
s_{b_k,-i}^t(-u)\ts\si_{b_1,\dots,\wh {b_k},\dots,b_{N-1}}(u-1)\ts\zeta
$$
$$
=\frac{2u+1}{2u\pm 1}\ts(N-2)!\ts\theta_{in}\sum_{k=1}^{N-1}
s_{i,-b_k}(-u)\ts(-1)^{k-1}\ts
\theta_{-b_k,n}\ts\si_{b_1,\dots,\wh {b_k},\dots,b_{N-1}}(u-1)\ts\zeta,
$$
where $(b_1,\dots,b_{N-1})=(-n,-n+1,\dots,n-1)$.
Put
$$
\si_{-b_k}(u-1):=(N-2)!\ts(-1)^{k-1}\ts
\theta_{-b_k,n}\ts\si_{b_1,\dots,\wh {b_k},\dots,b_{N-1}}(u-1).
$$
Then, taking into account (4.14), we get the following matrix
relation:
$$
\frac{2u+1}{2u\pm 1}\ts\wt S^{(n)}(-u)
\left( \matrix\si_{-n+1}(u-1)\\
\vdots\\
\si_n(u-1)
\endmatrix\right)=
\left( \matrix 0\\
\vdots\\
\wh s_{nn}(u)
\endmatrix\right)
$$
Multiplying its both sides by the matrix $(\wt S^{(n)}(-u))^{-1}$ from the
left and comparing the $n$-th coordinates of the vectors,
we obtain using (4.15) that
$$
\frac{2u+1}{2u\pm 1}\ts\sdet S^{(n-1)}(u-1)=
\bigl((\wt S^{(n)}(-u))^{-1}\bigr)_{nn}\wh s_{nn}(u),
$$
and hence,
$$
\wh s_{nn}(u)=\frac{2u+1}{2u\pm 1}\ts
|\wt S^{(n)}(-u)|_{nn}\ts\sdet S^{(n-1)}(u-1).
$$
Together with (4.11) this gives
$$
\sdet S(u)=\frac{2u+1}{2u\pm 1}\ts
|\wt S^{(n)}(-u)|_{nn}\cdot\sdet S^{(n-1)}(u-1)\cdot
|S(u-N+1)|_{nn}.\tag4.16
$$
Note that the series $|S(u-N+1)|_{nn}$
commutes with the matrix elements of the matrix $S^{(n-1)}(v)$
(see [MNO, Proposition 7.5]) and hence with the series
$\sdet S^{(n-1)}(u-1)$. Since
the coefficients of $\sdet S(u)$ belong to
the center of the algebra $\Y^{\pm}(N)$ all the factors in
the right hand side of (4.16) are mutually permutable.
So, (4.16) can be rewritten as
$$
\sdet S(u)=\frac{2u+1}{2u\pm 1}\ts
\sdet S^{(n-1)}(u-1)\cdot|\wt S^{(n)}(-u)|_{nn}\cdot
|S(u-N+1)|_{nn}.\tag4.17
$$
Performing an easy
calculation and using the induction argument we complete the proof
of formulae (4.7) and (4.8).

Further, using again the fact that the coefficients of the
Sklyanin determinant belong to the center of the Yangian, we obtain from
formula (4.17) that the product of the
quasi-determinants
$$
|\wt S^{(n)}(-u-N/2+1/2)|_{nn}\cdot|S^{(n)}(u-N/2+1/2)|_{nn}
$$
is permutable with each of the factors in
formula (4.7) or (4.8). On the other hand, it was noticed above that
the series $|S^{(n)}(u-N/2+1/2)|_{nn}$ is also permutable with
each of the factors, which proves the theorem.
\medskip

Theorem 2.2 can be derived now as follows.
It is not difficult to verify (see [O], [MNO]) that the mapping
$$
\xi: s_{ij}(u)\mapsto \delta_{ij}+F_{ij}(u\pm {1\over 2})^{-1}
$$
defines the algebra homomorphism
$$
\xi: \Y^{\pm}(N)\ra \U(\g(n)).
$$
Formula (4.6)
implies that
$$
\prod_{i=-n}^n(1+\rho_iu^{-1})\ts\xi(c(u))=
\wt{\det}\ts(1+(F+\rho)u^{-1}).
$$
On the other hand, if $1\leq m\leq n$ then in the case of
$N=2n$ we have
$$
(1+\rho_mu^{-1})\ts\xi(|S^{(m)}(u-m+1/2)|_{mm})=
|1+(F^{(m)}+\rho_m)u^{-1}|_{mm}
$$
and
$$
(1+\rho_{-m}u^{-1})\ts\xi(|\wt S^{(m)}(-u-m+1/2)|_{mm})=
|1-(\wt F^{(m)}+\rho_m)u^{-1}|_{mm},
$$
while in the case of $N=2n+1$
$$
(1+\rho_mu^{-1})\ts\xi(|S^{(m)}(u-m)|_{mm})=|1+(F^{(m)}+\rho_m)u^{-1}|_{mm}
$$
and
$$
(1+\rho_{-m}u^{-1})\ts\xi(|\wt S^{(m)}(-u-m)|_{mm})=
|1-(\wt F^{(m)}+\rho_m)u^{-1}|_{mm}.
$$
So, applying the homomorphism $\xi$ to both sides of the decomposition
(4.7) or (4.8) and replacing $u^{-1}$ with $t$ we complete the proof
of Theorem 2.2.

\bigskip
\noindent
{\bf Remarks.} (i) An analogue of
the Cayley--Hamilton theorem for a generic matrix $A$ with noncommutative
entries was obtained in [GKLLRT, Theorem 8.17] and the
{\it pseudo-determinants} $\det_iA$
of $A$ were introduced. In particular,
for $A=E$ or $F$ the Cayley--Hamilton identity
coincides with the characteristic identity satisfied by $E$ or $F$
(see resp. [NT] or [Mo1]) and the images of these identities in
highest weight representations give the Bracken--Green identities [BG],
[G].
The pseudo-determinants of the matrices $E+\rho_N$ or $F+\rho_n$
coincide with
the leading coefficients of the Capelli determinant $\det\ts(1-t(E+\rho))$
or the Capelli-type determinant $\wt{\det}\ts(1-t(F+\rho))$
respectively. In
particular, for any $i=-n,\dots,n$
$$
\text{\rm det}_i\ts(F+\rho_n)=(-1)^{N+n}
\sum_{p\in\Sym_N}\sgn(pp')
(F+\rho_{-n})_{-i_{p(1)},i_{p'(1)}}
\cdots(F+\rho_{n})_{-i_{p(N)},i_{p'(N)}}.
$$

(ii) The approach used in Section 1  for the
construction of Laplace operators for the Lie algebras $\gl(N)$,
$\oa(N)$, $\spa(N)$ can be applied to the algebras $\Y(N)$ and
$\Y^{\pm}(N)$ themselves. One can consider the specializations
of the symmetric functions taking the coefficients $d_1,d_2,\dots$ of
the quantum determinant or the coefficients $c_2,c_4,\dots$
of the Sklyanin determinant as elementary symmetric functions.
Then a description of other families of central elements of $\Y(N)$ and
$\Y^{\pm}(N)$ as complete or power sums symmetric functions can be obtained

by using Theorems 3.1 and 4.1.

\newpage
\heading
{\bf References}
\endheading
\bigskip

\itemitem{[BG]}
{A. J. Bracken and H. S. Green},
{\sl Vector operators and a polynomial identity for SO(n)},
{J. Math. Phys.}
{\bf 12}
(1971),
2099--2106.

\itemitem{[D]}
{V. G. Drinfeld},
{\sl Hopf algebras and the quantum Yang--Baxter equation}, {Soviet Math.
Dokl.}
{\bf 32}
(1985),
254--258.

\itemitem{[G]}
{H. S. Green},
{\sl Characteristic identities for generators of GL(n), O(n) and Sp(n)},
{J. Math. Phys.}
{\bf 12}
(1971),
2106--2113.

\itemitem{[GKLLRT]}
{I. M. Gelfand, D. Krob, A. Lascoux,
B. Leclerc, V. S. Retakh and J.-Y. Thibon},
{\sl Noncommutative symmetric functions},
{Preprint LITP 94.39,
Paris, 1994} (hep-th$\backslash9407124$)\newline
(to appear in Advances in Math.).

\itemitem{[GR1]} {I. M. Gelfand and V. S. Retakh}, {\sl Determinants
of matrices over noncommutative rings}, {Funct. Anal. Appl.} {\bf 25}
(1991), 91-102.

\itemitem{[GR2]} {I. M. Gelfand and V. S. Retakh}, {\sl A theory
of noncommutative determinants and characteristic functions of graphs},
{Funct. Anal. Appl.} {\bf 26} (1992), 1-20;  Publ. LACIM, UQAM, Montreal,
{\bf 14}, 1-26.

\itemitem{[H]}
{R. Howe},
{\sl Remarks on classical invariant theory}, {Trans. AMS}
{\bf 313}
(1989),
539--570.

\itemitem{[HU]}
{R. Howe and T. Umeda},
{\sl The Capelli identity, the double commutant theorem,
and multiplicity-free actions},
{Math. Ann.}
{\bf 290}
(1991),
569--619.

\itemitem{[KL]}
{D. Krob and B. Leclerc}, {\sl Minor identities for
quasi-determinants and quantum determinants}, Preprint LITP 93.46,
Paris, 1993.

\itemitem{[KS]}
{P. P. Kulish and E. K. Sklyanin},
{\sl Quantum spectral transform method: recent developments},
in \lq Integrable Quantum Field Theories', {Lecture Notes in Phys.}
{\bf 151}
Springer,
Berlin-Heidelberg,
1982,
pp. 61--119.

\itemitem{[M]}{I. G. Macdonald},
{\sl Symmetric functions and Hall polynomials},
{Oxford Math. Monographs}, Oxford University Press,
1979.

\itemitem{[Mo1]}
{A. I. Molev},
{\sl Yangians and classical Lie algebras, Part II. Sklyanin
determinant, Laplace operators and characteristic identities},
{Preprint CMA MRR 024-94, Canberra, 1994
(hep-th$\backslash9409036$)(to appear in J. Math. Phys.).

\itemitem{[Mo2]}
{A. I. Molev},
{\sl Yangians and Laplace operators
for classical Lie algebras},
Proceedings of the Conference
`Confronting the Infinite', Adelaide, February 1994 (to appear).

\itemitem{[MNO]}
{A. I. Molev, M. L. Nazarov and G. I. Olshanski\u\i},
{\sl Yangians and classical Lie algebras},
{Preprint CMA-MR53-93, Canberra}, 1993
(hep-th$\backslash9409025$).

\itemitem{[N]}
{M. L. Nazarov},
{\sl Quantum Berezinian and the classical Capelli identity},
{Lett. Math. Phys.}
{\bf 21}
(1991),
123--131.

\itemitem{[NT]}
{M. Nazarov and V. Tarasov},
{\sl Yangians and Gelfand--Zetlin bases},
Preprint RIMS, Kyoto Univ.,
February 1993 (to appear in Publications of RIMS, {\bf 30} (1994), no 3).

\itemitem{[O]}
{G. I. Olshanski\u\i},
{\sl Twisted Yangians and infinite-dimensional classical Lie algebras},
in \lq Quantum Groups (P. P. Kulish, Ed.)', {Lecture Notes in Math.}
{\bf 1510},
Springer,
Berlin-Heidelberg,
1992,
pp. 103--120.

\itemitem{[PP]}
{A. M. Perelomov and V. S. Popov},
{\sl Casimir operators for semi-simple Lie algebras}, {Isv. AN SSSR}
{\bf 32}
(1968),
1368--1390.

\itemitem{[S]}{R. P. Stanley},
{\sl Enumerative combinatorics, I},
Wadsworth and Brooks/Cole, Monterey, California
1986.

\itemitem{[TF]}
{L. A. Takhtajan and L. D. Faddeev},
{\sl Quantum inverse scattering method and the Heisenberg
$XYZ$-model},
{Russian Math. Surv.}
{\bf 34}
(1979),
no. 5,
11--68.

\enddocument
\end